\documentclass[conference,letterpaper]{IEEEtran}

\usepackage[utf8]{inputenc} 
\usepackage[T1]{fontenc}
\usepackage{url}
\usepackage{ifthen}
\usepackage{cite}
\usepackage[cmex10]{amsmath} 
\usepackage{mathtools}
\usepackage{graphicx}
\usepackage{amsfonts}
\usepackage{enumitem}
\usepackage{amssymb}
\usepackage[]{hyperref}
\usepackage{cleveref}

\DeclarePairedDelimiter\floor{\lfloor}{\rfloor}

\makeatletter
\def\@begintheorem#1#2{\par\vspace{0.4\baselineskip}\noindent\textbf{#1~#2.} \itshape}
\def\@opargbegintheorem#1#2#3{\par\vspace{0.4\baselineskip}\noindent\textbf{#1~#2~(#3).} \itshape}
\makeatother

\newtheorem{thm}{Theorem}

\newtheorem{lem}[thm]{Lemma}

\newtheorem{rem}[thm]{Remark}

\newenvironment{casenv}
{\begin{enumerate}[label=\Roman*., ref=\Roman*]}
{\end{enumerate}}

\newenvironment{customproof}[1]{%
    \par\vspace{\baselineskip}\textit{Proof of #1:}\ }{\hfill$\blacksquare$\par}

\AtBeginDocument{\providecommand\figref[1]{Fig.~\ref{fig:#1}}}
\AtBeginDocument{}
\AtBeginDocument{\providecommand\lemref[1]{Lemma \ref{lem:#1}}}
\AtBeginDocument{\providecommand\thmref[1]{Theorem \ref{thm:#1}}}
\AtBeginDocument{}
\AtBeginDocument{}

\newcommand{\Ex}[1]{\left\langle#1\right\rangle}

\newcommand{\px}{p_X}
\newcommand{\py}{p_Y}

\begin{document}
\title{Capacity-Achieving Input Distribution of the Additive Uniform Noise Channel With Peak Amplitude and Cost Constraint} 

\author{%
 \IEEEauthorblockN{Jonas Stapmanns, Catarina Dias, Luke Eilers and Jean-Pascal Pfister}
\IEEEauthorblockA{Department of Physiology \\
                     University of Bern\\
                     Bern, Switzerland\\
                   Email: \{jonas.stapmanns, catarina.reisdias, luke.eilers, jeanpascal.pfister\}@unibe.ch} }

\maketitle

\begin{abstract}
   Under which condition is quantization optimal? We address this question in the context of the additive uniform noise channel under peak amplitude and power constraints. We compute analytically the capacity-achieving input distribution as a function of the noise level, the average power constraint and the exponent of the power constraint. We found that when the cost constraint is tight and the cost function is concave, the capacity-achieving input distribution is discrete, whereas when the cost function is convex, the support of the capacity-achieving input distribution spans the entire interval. 
\end{abstract}

\section{Introduction}
Since Shanon introduced channel capacity\cite{shannon_mathematical_1948}, capacity-achieving input distributions have been studied for several combinations of channels and constraints. \cite{shannon_mathematical_1948,smith_information_1971,oettli_capacity-achieving_1974,shamai_capacity_1995,lapidoth_capacity_2009,dytso_when_2018,dytso_capacity_2019,dytso_capacity_2020,eisen_capacity-achieving_2023,barletta_binomial_2024}. In the absence of peak amplitude constraint (PA), some channels (such as the additive Gaussian channel with variance constraint) have a capacity-achieving distribution with continuous support, while in the presence of PA it has been shown that some channels (such as the additive Gaussian channel \cite{smith_information_1971}, the Poisson channel \cite{lapidoth_capacity_2009} or the additive channel with piecewise linear noise \cite{oettli_capacity-achieving_1974}) have a discrete capacity-achieving input distribution, i.e. all the input is concentrated on a finite set of mass points. 

So what are the necessary and sufficient conditions such that the capacity-achieving input distribution is discrete? Is there a channel for which there is a phase transition between a continuous capacity-achieving input distribution and a discrete one? Because of its analytical tractability, we frame those questions in the context of the additive uniform noise channel with PA and power constraint. We found that when the cost constraint is tight and concave, the capacity-achieving input distribution is discrete, whereas it has continuous support when the cost function is convex.

\section{Problem statement}

We investigate the capacity-achieving input distribution of the additive
channel 
\begin{equation}
Y=X+N,\quad\mathrm{where}\,N\sim\mathrm{Uniform}\left(-b,b\right),\label{eq:cahnnel_definition}
\end{equation}
with $b>0$. Hence, the density of the noise is given by 
$p_{N}\left(y\mid x\right)=\mathbf{1}_{x-b<y<x+b}/\left(2b\right)$.
For convenience, we define an additional variable for the inverse
width $r\coloneqq1/\left(2b\right)$. The input to the channel is
subject to the PA $P\left(X<0\right)=P\left(X>1\right)=0$
and, additionally, to the cost constraint 
\begin{equation}
    \Ex{c(x)}\leq\bar{c}, \quad\mathrm{with}\quad 
    c(x) = x^\alpha,~\alpha > 0 
\end{equation}
Unless specified otherwise, the expectation $\Ex{\cdot}$ is w.r.t to the input distribution that will be denoted as $\px$.  

\section{Results}

In \cite{smith_information_1971}, Smith derived necessary and sufficient conditions
for $\px^{\ast}$ to be the capacity-achieving input distribution
of a channel with additive noise and PA. Even though he considers
Gaussian additive noise and a constraint on the second moment, i.e.
$c\left(x\right)=x^{2}$, his derivation of the following lemma holds
for arbitrary additive noise and arbitrary cost function.

\begin{thm}
(Optimality conditions; Smith, \cite{smith_information_1971}) Let $C$ denote the channel capacity. Then, for an additive
channel with PA and a cost constraint of the form $\left\langle c\right\rangle \leq \bar{c}$,
the capacity-achieving input distribution $\px^{\ast}$ implicitly
defined by
\begin{equation}
C=\max_{\begin{array}{c}
\px \colon \\
\int_0^1 dx \, \px(x) = 1\\
\left\langle c\left(x\right)\right\rangle \leq\bar{c}
\end{array}}I\left(X;Y\right),\label{eq:channel_capacity}
\end{equation}
is unique and determined by the necessary and sufficient conditions
\begin{alignat}{2}
i\left(x;\px^{\ast}\right) & \leq I\left(\px^{\ast}\right)+\lambda\left(c\left(x\right)-\bar{c}\right) & \quad & \mathrm{for}\;\mathrm{all}\;x\in\left[0,1\right],\label{eq:ineq_constr}\\
i\left(x;\px^{\ast}\right) & =I\left(\px^{\ast}\right)+\lambda\left(c\left(x\right)-\bar{c}\right) & \quad & \mathrm{for}\;\mathrm{all}\;x\in S,\label{eq:eq_constr}
\end{alignat}
where $S$ denotes the support of $\px$,
\begin{equation}
i\left(x;\px\right) \coloneqq\int dy\,p_{\mathrm{N}}\left(y\mid x\right)\log\frac{p_{\mathrm{N}}\left(y\mid x\right)}{\py\left(y;\px\right)}
\end{equation}
 is the marginal information density, and $I\left(\px\right)\coloneqq\int dx\,\px\left(x\right)\,i\left(x;\px\right)$
is the mutual information between $X$ and $Y$. In the absence of the cost constraint, or if the constraint is not tight, it holds $\lambda=0$ and $\px^{0\ast}$ denotes the corresponding capacity-achieving input distribution.
\end{thm}
\begin{IEEEproof}
See \cite{smith_information_1971}, replacing the variance constraint $x^{2}$ by the general cost constraint $c\left(x\right)$.
\end{IEEEproof}
%

\begin{figure}
\includegraphics[width=1.0\columnwidth]{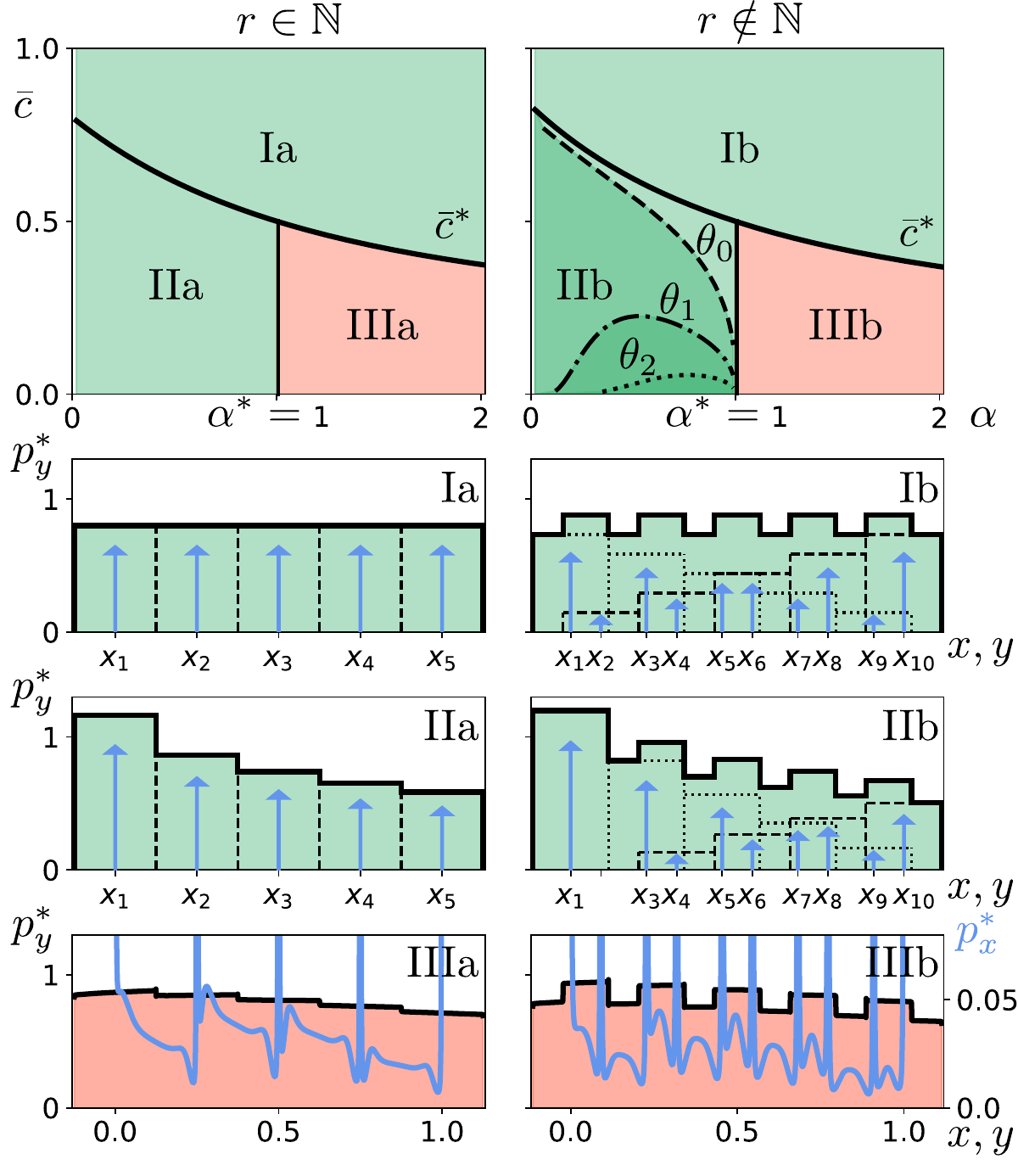}
\caption{The different cases discussed in \thmref{main}.
In the left column $r\in\mathbb{N}$ ($r=4$) and in the right column $r\notin\mathbb{N}$ ($r=4.4$).
Top: Phase diagram in the $\alpha$-$\bar{c}$-plane.
Green and red background indicate $\px^{\ast}$ with discrete support and
support on the entire interval $[0,1]$, respectively.
Ia,b and IIa,b: discrete $\px^{\ast}$ with masses and positions indicated by the heights and the positions
of the blue arrows, corresponding $p_N(y\mid x)$ by dashed boxes in Ia/IIa and by dotted (j odd)
and dashed (j even) boxes in Ib/IIb.
The black line is the resulting $\py^{\ast}$.
IIIa and IIIb: numerical result for $\px^{\ast}$ (blue) using the Blahut-Arimoto algorithm \cite{blahut_1972, arimoto_1972} and corresponding $\py^{\ast}$ in black.\vspace{-0.3cm}
\label{fig:sketch_integer_case}}
\end{figure}

For given values of $\alpha$ and $r$, we define the critical expected cost $\bar{c}^\ast \coloneqq \Ex{c(x)}_{\px^{0\ast}}$ as the cost below which the cost constraint becomes tight.
\begin{thm}
\label{thm:main}(Main Theorem) The capacity-achieving input distribution $\px^{*}$
of the additive uniform noise channel with peak amplitude and cost
constraint with cost function $c(x) = x^\alpha$, $\alpha>0$, has the following properties:
\begin{casenv}
\item[I] (Oettli, \cite{oettli_capacity-achieving_1974}) \label{case:i}If the cost constraint is inactive (i.e. $\bar{c}\geq\bar{c}^{\ast}$),
then the capacity-achieving input distribution is given by $\px^{*}=\sum_{j=1}^{N_r}m_{j}\delta(x-x_{j})$ where 
\begin{align}
N_r & =\begin{cases}
n+1 & \mathrm{if}\;r\in\mathbb{N}\\
2n+2& \mathrm{if}\;r\notin\mathbb{N}
\end{cases}\label{eq:npoints}
\end{align}
is the number of mass points with $n\coloneqq\floor{r}$. The mass locations $x_{j}$ and the masses $m_{j}$ are given
by
\begin{align}
x_{j} & =\begin{cases}
\frac{j-1}{n} & \mathrm{if}\;r\in\mathbb{N}\\
\frac{j-1}{2r} & \mathrm{if}\;r\notin\mathbb{N},\;j\;\mathrm{is}\;\mathrm{odd}\\
1-\frac{2n+2-j}{2r} & \mathrm{if}\;r\notin\mathbb{N},\;j\;\mathrm{is}\;\mathrm{even},
\end{cases}\label{eq:def_pos_unconstr}\\
m_{j} & =\begin{cases}
\frac{1}{n+1} & \mathrm{if}\;r\in\mathbb{N}\\
\frac{2n+2-\left(j-1\right)}{2\left(n+1\right)\left(n+2\right)} & \mathrm{if}\;r\notin\mathbb{N},\;j\;\mathrm{is}\;\mathrm{odd}\\
\frac{j}{2\left(n+1\right)\left(n+2\right)} & \mathrm{if}\;r\notin\mathbb{N},\;j\;\mathrm{is}\;\mathrm{even},
\end{cases}\label{eq:def_masses_unconstr}
\end{align}
where $j=1,\ldots,N_r$. Thus, the support
of $\px^{\ast}$ is discrete and given by $S_{0}\coloneqq\left\{ x_{j}\mid j=1,\ldots,N_r\right\} $.
\figref{sketch_integer_case} (Ia and Ib) illustrate $\px^\ast$.
\item[IIa] \label{case:iia}If the cost constraint is active ($\bar{c}<\bar{c}^{\ast}$), the
cost function $c(x)$ is concave ($\alpha\leq1$), and $r\in\mathbb{N}$, then the capacity-achieving input distribution is discrete with mass locations
as in (\ref{eq:def_pos_unconstr}) and masses given by
\begin{equation}
m_{j} =\frac{1}{z}e^{-\lambda^\ast c_{j}}, \quad z =\sum_{j=1}^{N_r}e^{-\lambda^\ast c_{j}},\label{eq:masses_integer_case}
\end{equation}
for some $\lambda^\ast > 0$ and with $c_{j}=x_{j}^{\alpha}$.
Thus, the support of $\px^{\ast}$is given by $S_{0}$,
see \figref{sketch_integer_case} (IIa) for an illustration of $\px^\ast$.
\item[IIb] \label{case:iib} If the cost constraint is active ($\bar{c}<\bar{c}^{\ast}$), the
cost function $c(x)$ is concave ($\alpha\leq1$), and $r\notin\mathbb{N}$, then
the capacity-achieving input distribution is discrete. Furthermore, there exist $n$ thresholds $0<\theta_{n-1}<\dots<\theta_0<\bar{c}^*$ such that the support can be expressed as
\begin{align}
    S  = \begin{cases}
S_0 & \mathrm{if} ~ \bar{c} >\theta_0\\
S_k & \mathrm{if} ~ \bar{c}\in (\theta_k,\theta_{k-1}],~1\leq k\leq n-1\\
S_n & \mathrm{if} ~ \bar{c}\in (0,\theta_{n-1}]
\end{cases},\label{eq:Sk_cases}
 \end{align}
where $S_k = S_{k-1}\setminus \{x_{2k}\}$, $1\leq k\leq n$.
\figref{sketch_integer_case} (IIb) illustrates $p_x^\ast$ for
$\bar{c}\in(\theta_1,\theta_0]$.
\item[III] \label{case:iii}If the cost constraint is active ($\bar{c}<\bar{c}^{\ast}$) and
the cost function is $c(x)$ convex ($\alpha>1$), then the capacity
achieving input distribution has support on the entire interval $[0,1]$,
see \figref{sketch_integer_case} (IIIa and IIIb).
\end{casenv}
\end{thm}
The top row of \figref{sketch_integer_case} shows the positions of the different cases
of \thmref{main} in the phase diagram.

\begin{customproof}{Case I (Oettli)}
The full proof is given in \cite{oettli_capacity-achieving_1974}. 
The idea of the proof is to show that the resulting
$p_{0,y}^{\ast}\left(y;p_{0,x}^{\ast}\right)$ is $2b$-periodic within
the interval $D_{Y}\coloneqq\left[-b,1+b\right]$, which leads to a constant
$i(x;\px^\ast)=I(X;Y)$ and therefore the
necessary and sufficient conditions (\ref{eq:ineq_constr}) and (\ref{eq:eq_constr})
(with $\lambda=0$) are fulfilled with equality in (\ref{eq:ineq_constr}).
\end{customproof}
\begin{rem}
\label{rem:r_in_N_consistency}Note that if $r\in\mathbb{N}$,
i.e. $r=n$, the width of the blocks $p_{N}\left(y\mid x\right)$
is such that a number $n$ of these blocks can cover the interval
$D_{Y}$ perfectly without overlaps or gaps. Letting $\rho:=r-n$ approach
$0$ from above, $\lim_{\rho\searrow 0}x_{j}=\lim_{\rho\searrow 0}x_{j+1}=\left(j-1\right)/\left(2n\right)$,
for $j=1,3,\ldots,N_{r}-1$ odd, and their masses add up to $1/\left(n+1\right)$.
In this configuration, the touching blocks
form a uniform output distribution $p_{0,y}^{\ast}\left(y\right)=\frac{n}{n+1}\mathbf{1}_{-b<y<1+b}$, see \figref{sketch_integer_case} (Ia),
which is known to maximize the output entropy $H\left(Y\right)\coloneqq-\int dy\,\int dx\,p_{N}\left(y\mid x\right)\log \py\left(y\right)$
if $Y$ is restricted to an interval $D_{Y}$.
\end{rem}

\begin{customproof}{Case IIa}
For $\alpha\leq1$ and $r\in\mathbb{N}$, we will show that the input distribution
$\px^{*}=\sum_{j=1}^{N_r}m_{j}\delta(x-x_{j})$ with $x_j$ and $m_j$ as defined in
(\ref{eq:def_pos_unconstr}) and (\ref{eq:def_masses_unconstr}) fulfills the necessary
and sufficient conditions (\ref{eq:ineq_constr}) and (\ref{eq:eq_constr}).

The positions $x_j$ are such that their outputs, including the noise
$p_{\mathrm{N}}\left(y\mid x_{j}\right)$, cover
the y-axis without overlap or gaps within the interval $D_{Y}$. The
corresponding marginal information density is given by
$i\left(x_{j};\px^{\ast}\right)=-\log m_{j}$, so
that the equality constraint (\ref{eq:eq_constr}) evaluates to
\begin{equation}
-\log m_{j}=I+\lambda\left(c_{j}-\bar{c}\right),\;j=1,\ldots,N_r,\label{eq:eq_constr_integer_case}
\end{equation}
where $c_{j}\coloneqq c\left(x_{j}\right)$. The masses $m_{j}=m_{j}\left(\lambda\right)$,
and hence the corresponding probability distribution $\px^{\lambda}$,
depend on $\lambda$, but this dependence is omitted when clear from
context. Computing the difference between two consecutive $j$ yields
$n$ equations of the form
\begin{equation}
m_{j+1}=m_{j}e^{-\lambda\left(c_{j+1}-c_{j}\right)}.\label{eq:diff_eqs_integer_case}
\end{equation}
The $m_{j}$ are nonnegative and a decreasing series over $j$ because
$\lambda\geq0$ and $c_{j+1}-c_{j}>0$. Since $\sum_{j=1}^{N_r}m_{j}=1$,
the masses can be written in the form of (\ref{eq:masses_integer_case}).
With the following lemma, we prove that a unique $\lambda^\ast$ fulfills the cost constraint.
\begin{lem}
\label{lem:chain_non_overlapping}(Uniqueness of $\lambda^\ast$, $r\in\mathbb{N}$)
For fixed $r\in\mathbb{N}$ and any given $\bar{c}\in\left(0,\bar{c}^{\ast}\right]$,
there is a unique solution to the equality constraint (\ref{eq:eq_constr}) given by
the discrete probability distribution $\px^{\lambda^{\ast}}\left(x\right)=\sum_{j=1}^{N_r}m_{j}\left(\lambda^{\ast}\right)\delta\left(x-x_{j}\right)$,
with $m_{j}\left(\lambda\right)$ and $x_j$ as defined in (\ref{eq:def_pos_unconstr}) and (\ref{eq:def_masses_unconstr}).
\end{lem}
\begin{IEEEproof}
By construction, for a given $\lambda$, the $n+1$ masses
$m_{j}$ fulfill the $n$ difference equations (\ref{eq:diff_eqs_integer_case})
and the corresponding cost is given by $\left\langle c\left(x\right)\right\rangle _{\px^{\lambda}}$.
This is equivalent to the $n+1$ original equations (\ref{eq:eq_constr_integer_case})
with $\bar{c}=\left\langle c\left(x\right)\right\rangle _{\px^{\lambda}}$. 
When $\lambda=0$, the constraint is inactive and $m_{j}\left(0\right)=1/\left(n+1\right)$,
so that $\left\langle c\left(x\right)\right\rangle _{\px^{0\ast}}=\bar{c}^{\ast}$.
In the opposite limit of $\lambda\rightarrow\infty$, all the probability
is concentrated at zero, i.e. $\lim_{\lambda\rightarrow\infty}m_{1}\left(\lambda\right)=1$,
and for all other masses, $j>1$, $\lim_{\lambda\rightarrow\infty}m_{j}\left(\lambda\right)=0$,
which yields $\lim_{\lambda\rightarrow\infty}\left\langle c\left(x\right)\right\rangle \left(\lambda\right)=0$.
In between the two extremes, $\left\langle c\left(x\right)\right\rangle \left(\lambda\right)$
is a strictly monotonic decreasing function. Defining $c_{j}\coloneqq c\left(x_{j}\right)$,
we obtain
\begin{align}
&\frac{\partial}{\partial\lambda}\left\langle c\left(x\right)\right\rangle _{\px^{\lambda}} =\frac{\partial}{\partial\lambda}\sum_{j=1}^{N_r}m_{j}\left(\lambda\right)\,c_{j}
 =\frac{\partial}{\partial\lambda}\sum_{j} c_{j}\frac{e^{-\lambda c_{j}}}{z\left(\lambda\right)}\nonumber\\
 =&\sum_{j} c_{j}\frac{-c_{j}e^{-\lambda c_{j}}z\left(\lambda\right)+e^{-\lambda c_{j}}\sum_{k}c_{k}e^{-\lambda c_{k}}}{z^{2}\left(\lambda\right)}\nonumber\\
 =&-\sum_{j} c_{j}^{2} \, m_{j}\left(\lambda\right)
 +\biggl(\sum_{j}m_{j}\left(\lambda\right)\,c_{j}\biggr)\biggl(\sum_{k}m_{k}\left(\lambda\right)\,c_{k}\biggr)\nonumber\\
 =&-\left(\left\langle c^{2}\left(x\right)\right\rangle _{\px^{\lambda}}-\left\langle c\left(x\right)\right\rangle _{\px^{\lambda}}^{2}\right)
 =-\mathrm{Var}_{\px^{\lambda}}\left(c\right)\leq0,
\end{align}
with equality if and only if the total mass is concentrated on $m_{1}$,
i.e. in the case $\lambda\rightarrow\infty$. Therefore, if $\bar{c}\in(0,\bar{c}^\ast]$ there is one unique
$\px^{\lambda^{\ast}}$ such that $\left\langle c\left(x\right)\right\rangle _{\px^{\lambda^{\ast}}}=\bar{c}$.
\end{IEEEproof}\vspace{0.4\baselineskip}
To show that probability distributions with support $S_0$ satisfy the inequality constraint (\ref{eq:ineq_constr}), we use the following lemma.
\begin{lem}
\label{lem:linear_i}(Piece-wise linearity of the marginal information density)
If the positions $x_{j}$ are defined as in (\ref{eq:def_pos_unconstr}),
and the corresponding masses are nonnegative,  $m_{j}\geq0$, then $i\left(x;\px\right)$
is linear for $x\in\left[x_{j},x_{j+1}\right]$ with slope $r\log\left[\left(m_{j-1}+m_{j}\right)/\left(m_{j+1}+m_{j+2}\right)\right]$
if $m_{j-1}+m_{j}\neq0$ $\forall j=2,\dots,N_r-2$. 
\end{lem}
\begin{IEEEproof}
We consider the case $r\notin \mathbb{N}$. The case $r \in \mathbb{N}$ follows as a special case. For $x\in\left[x_{j},x_{j+1}\right]$, $\py\left(y\right)$ consists
of three piece-wise constant segments between the positions $x-b \leq x_{j+1}-b\leq x_j+b \leq x+b$ (cf. Figure \ref{fig:sketch_integer_case} Ib).
With $d\coloneqq x-x_{j}$, the marginal information density 
evaluates to
\begin{align}
i\left(x;\px\right) & =-\frac{1}{2b}\int_{x-b}^{x+b}dy\,\log\left[2b\,\py^{\ast}\left(y\right)\right]\nonumber\\
 & =-r\left(x_{j+1}-x_j -d\right)\log\left(m_{j-1}+m_{j}\right)\nonumber\\
 & \hphantom{=}-r\left(2b-x_{j+1}+x_j\right)\log\left(m_{j}+m_{j+1}\right)\nonumber\\
 & \hphantom{=}-r\,d\log\left(m_{j+1}+m_{j+2}\right)\nonumber\\
 & =r\log\left(\frac{m_{j-1}+m_{j}}{m_{j+1}+m_{j+2}}\right)d+D,
\end{align}
where all terms independent of $d$ are absorbed into $D$.
\end{IEEEproof}
\begin{rem}
This linear interpolation of $i\left(x;\px\right)$ between two
consecutive $x_{j}$ is true for all $j^{\prime}=1,\ldots,2n+1$.
For $j^{\prime}=1$ and $j^{\prime}=2n+1$ one can set $m_{0}=0$
or $m_{2n+3}=0$, respectively, in the proof. When $r\in\mathbb{N}$,
we can combine the masses $m_{j-1}+m_{j}\rightarrow m_{j/2}$ for any even $j$.
\end{rem}
\begin{figure}
\includegraphics[width=1\columnwidth]{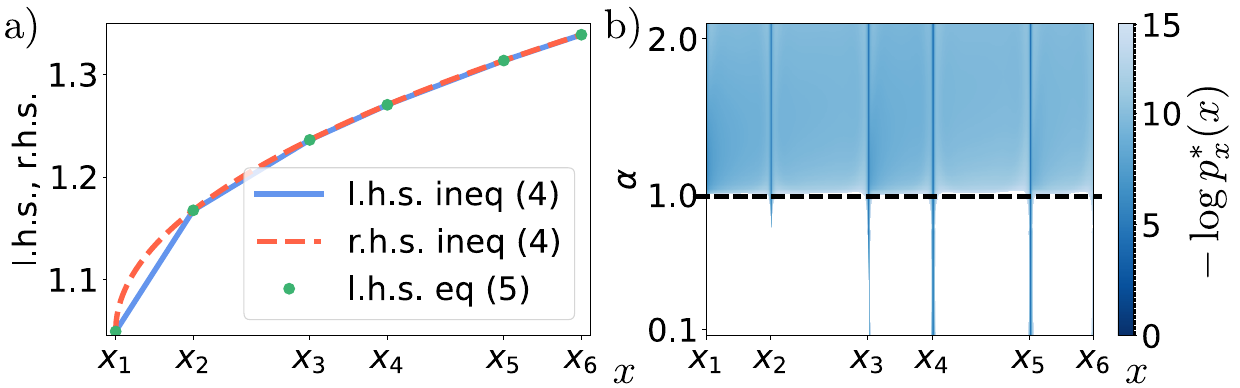}
\caption{a) The r.h.s. and the l.h.s. of (\ref{eq:ineq_constr}), illustrating
the linear interpolation between the points of support, where (\ref{eq:eq_constr})
ensures equality.  Other parameters: $r=2.4$ and $\bar{c}=0.54<\bar{c}^\ast$. b) $\px^\ast\left(x\right)$ as a function of $\alpha$ obtained
numerically by means of the Blahut-Arimoto algorithm \cite{blahut_1972,arimoto_1972}.
For $\alpha\protect\leq1$, $\px$ is discrete and for $\alpha>1$, it has support on
the entire interval $[0,1]$. Other parameters: $r=2.4$ and $\bar{c}=0.35<\bar{c}^\ast$. \vspace{-0.3cm}
\label{fig:cost}}
\end{figure}
\vspace{0.4\baselineskip}

To conclude the proof of Case IIa, we note that for every $\bar{c}\in\left[0,\bar{c}^{\ast}\right]$ 
\lemref{chain_non_overlapping} guarantees the
existence of a unique $\px^{\lambda^{\ast}}$ that solves (\ref{eq:eq_constr}).
Therefore, (\ref{eq:ineq_constr}) is satisfied with equality at $x_{j}$,
$j=1,\ldots,N_r$. Moreover, on the l.h.s. of (\ref{eq:ineq_constr}),
$i\left(x;\px^{\ast}\right)$ increases linearly between $x_{j}$
and $x_{j+1}$ because $m_{j}>m_{j+1}$, and the r.h.s is concave due
to $\alpha\leq1$. Thus, (\ref{eq:ineq_constr}) is also satisfied
for all the points $x\in\left(x_{j},x_{j+1}\right)$. Hence, $\px^{\lambda^{\ast}}$
is the capacity-achieving input distribution $\px^{\ast}$ and its
support is $S_0$, i.e. that of the unconstrained case.
This proves Case IIa. 
\end{customproof}
\begin{figure}
\includegraphics[width=1.0\columnwidth]{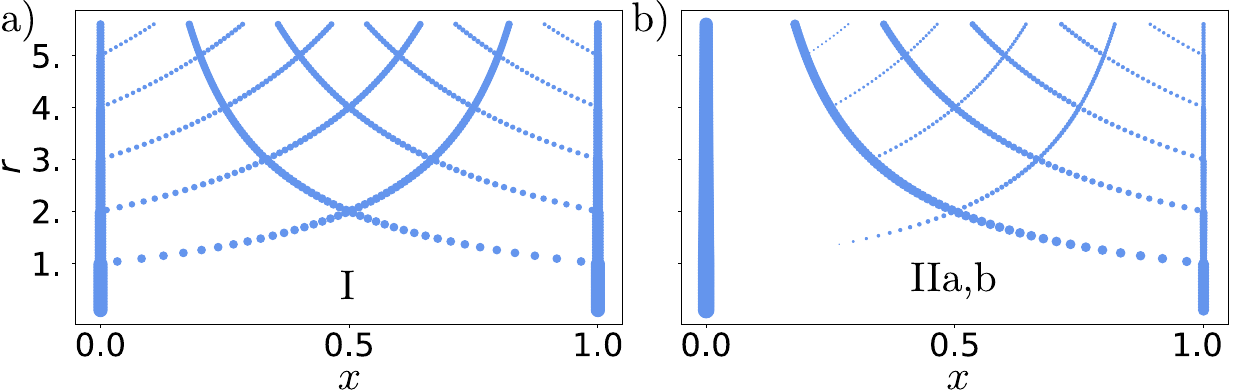}
\caption{Capacity-achieving input distribution $\px^{\ast}\left(x\right)$
as a function of $r$ for inactive (panel a)) and tight (panel b)) cost constraint. The diameter of the dots represents
the mass. Other parameters: $\bar{c}=3$ in b) and $\alpha=0.7$ in both.\vspace{-0.3cm}
\label{fig:capacity_achieving_input}}
\end{figure}
\begin{customproof}{Case IIb}
In the non-integer case, i.e. $r\notin\mathbb{N}$, we proceed in three steps corresponding
to the three cases in (\ref{eq:Sk_cases}). First, in \emph{Step A}, we focus on
$\bar{c} > \theta_0$,
where the support is given by $S_0$. Similarly to our proof of Case IIa, 
we derive the form of the capacity-achieving distribution and show that it
satisfies the necessary and sufficient conditions (\ref{eq:ineq_constr}) and (\ref{eq:eq_constr}).
Then, in \emph{Step B}, we briefly sketch the steps needed to prove iteratively that 
if the support is given by $S_{k-1}$ when $\bar{c}\in(\theta_{k-1},\theta_{k-2}]$, then the support is $S_{k}$ when $\bar{c}\in(\theta_{k},\theta_{k-1}]$. 
Finally, \emph{in Step C}, we show that $S_n$ is the support in the remaining interval $\bar{c}\in (0,\theta_{n-1}]$. The proofs of Lemmas \ref{lem:smallest-weight}, \ref{lem:chain_composed} and \ref{lem:ineq_at_x_2k} will be provided in the extended version of this manuscript.


To prove \emph{Step A} (i.e. for $\bar{c} > \theta_0$), we assume that the positions of the masses are given by (\ref{eq:def_pos_unconstr}) and show \emph{a posteriori} that those positions are optimal. Using $\rho =  r-n$, the marginal information density evaluates to
\begin{equation}
i\left(x_{j},\px\right) =-\rho\log\hat{m}_{\hat{f}\left(j\right)}-\left(1-\rho\right)\log\bar{m}_{\bar{f}\left(j\right)},\label{eq:i_non_integer}
\end{equation}
where the labels are given by $\hat{f}\left(j\right) \!=\!\floor{j/2}+1$ and $\bar{f}\left(j\right) \!=\!\floor{\left(j+1\right)/2}$,
with $j\!=\!1,\ldots,N_r$, and $\hat{m}$ and $\bar{m}$ are defined as
\begin{align}
\hat{m} & \coloneqq\left(m_{1},m_{2}+m_{3},\ldots,m_{2n}+m_{2n+1},m_{2n+2}\right),\label{eq:def_m_hat}\\
\bar{m} & \coloneqq(m_{1}+m_{2},m_{3}+m_{4},\ldots,m_{2n+1}+m_{2n+2}).\label{eq:def_m_bar}
\end{align}
Their entries correspond to the overlaps of $p_{\mathrm{N}}\left(y\mid x_{j}\right)$
and $p_{\mathrm{N}}\left(y\mid x_{j+1}\right)$, and their sums equal
the sum of all masses, i.e. $\sum_{j=1}^{n+2}\hat{m}_{j}=\sum_{j=1}^{n+1}\bar{m}_{j}=\sum_{j=1}^{N_r}m_{j} =1 $.
Inserting (\ref{eq:i_non_integer}) into (\ref{eq:eq_constr}),
subtracting the $\left(2j\right)$-th equality of the equality constraint
from the $\left(2i-1\right)$-th equality, and subtracting the $\left(2j+1\right)$-th
from the $\left(2j\right)$-th equality gives
\begin{align}
\hat{m}_{j+1} & =\hat{m}_{j}e^{-\lambda\frac{\widehat{\Delta c}_{j+1}}{\rho}},\;j=1,\dots,n+1,\\
\bar{m}_{j+1} & =\bar{m}_{j}e^{-\lambda\frac{\overline{\Delta c}_{j+1}}{1-\rho}},\;j=1,\dots,n,
\end{align}
respectively. Here, we defined 
\begin{align}
\widehat{\Delta c} & \coloneqq\left(0,c_{2}-c_{1},c_{4}-c_{3},\ldots,c_{2n+2}-c_{2n+1}\right),\\
\overline{\Delta c} & \coloneqq\left(0,c_{3}-c_{2},c_{5}-c_{4},\ldots,c_{2n+1}-c_{2n}\right).
\end{align}
\!
The masses $\hat{m}$ and $\bar{m}$, and hence the corresponding probability
distribution $\px^{\lambda}$, depend on $\lambda$ but this dependence
is omitted when clear from context. Including the sum constraint $\sum_{j=1}^{N_r}m_{j}=1$,
we can write
\begin{align}
\hat{m}_{j} & =\frac{1}{\hat{z}}e^{-\frac{\lambda}{\rho}\sum_{i=1}^{j}\widehat{\Delta c}_{i}}, & \hat{z}= & \sum_{j=1}^{n+2}e^{-\frac{\lambda}{\rho}\sum_{i=1}^{j}\widehat{\Delta c}_{i}}\label{eq:m_hat_j_exlp-1}\\
\bar{m}_{j} & =\frac{1}{\bar{z}}e^{-\frac{\lambda}{1-\rho}\sum_{i=1}^{j}\overline{\Delta c}_{i}}, & \bar{z}= & \sum_{j=1}^{n+1}e^{-\frac{\lambda}{1-\rho}\sum_{i=1}^{j}\overline{\Delta c}_{i}}.\label{eq:m_bar_j_expl-1}
\end{align}
Using (\ref{eq:def_m_hat}) and (\ref{eq:def_m_bar}), we can transform
back from $\hat{m}$ and $\bar{m}$ to the original masses
\begin{align}
m_{j} & =\begin{cases}
\sum_{k=1}^{\left(j+1\right)/2}\hat{m}_{k}-\sum_{k=1}^{\left(j-1\right)/2}\bar{m}_{k}, & j\;\mathrm{odd}\\
\sum_{k=1}^{j/2}\left(\bar{m}_{k}-\hat{m}_{k}\right), & j\;\mathrm{even}
\end{cases},\label{eq:def_masses_non_integer_case}
\end{align}
for $j=1,\ldots,N_r$. However, a priori it is not guaranteed that
$m_{j}>0$ for all $j$ independent of $\lambda$. In the special
case $\lambda=0$, we obtain the masses (\ref{eq:def_masses_unconstr})
of the unconstraint case I, where $m_{j}>0$ for all $j$. With the
following lemma, we show that a solution with only positive weights exists also for increasing $\lambda>0$.
\begin{lem}
\label{lem:smallest-weight}($m_2$ vanishes first) For $\alpha\leq 1$, $\rho>0$ and $\bar{c}\in (\theta_0,\bar{c}^\ast]$, there exists $\lambda_{0}>0$ such that for
every $\lambda\in[0,\lambda_{0})$, the masses defined by (\ref{eq:def_masses_non_integer_case})
satisfy $0<m_{2}<m_{j\neq2}$, $j=1,3,\ldots,N_r$. When $\lambda = \lambda_0$, the second mass vanishes, i.e. $0=m_{2}<m_{j\neq2}$, $j=1,3,\ldots,N_r$.
\end{lem}

\lemref{smallest-weight} ensures that $\px^{\lambda}\left(x\right)$,
$\lambda\in[0,\lambda_{0})$ is a valid probability distribution.
The following lemma proves the uniqueness of $\px^{\lambda}$ for
a given cost constraint $\bar{c}\in(\theta_{0},\bar{c}^{\ast}]$.
\begin{lem}
\label{lem:chain_overlapping}(Uniqueness of $\lambda^{\ast}$, $r\notin\mathbb{N}$)
For $\alpha\leq 1$, $r\notin\mathbb{N}$ and $\bar{c}\in(\theta_{0},\bar{c}^{\ast}]$,
there is a unique solution to the equality constraint (\ref{eq:eq_constr})
given by the discrete probability distribution $\px^{\lambda^\ast}\left(x\right)=\sum_{j=1}^{N_r}m_{j}\left(\lambda^{\ast}\right)\,\delta\left(x-x_{j}\right)$,
with $m_{j}\left(\lambda^\ast\right)$ and $x_{j}$ as defined in Eqs. (\ref{eq:def_masses_non_integer_case})
and (\ref{eq:def_pos_unconstr}).
\end{lem}
\begin{IEEEproof}
By construction, for a given $\lambda$, the $N_r$ masses $m_{j}$
fulfill the $N_r-1$ difference equations (\ref{eq:diff_eqs_integer_case}),
and the corresponding costs are given by $\left\langle c\left(x\right)\right\rangle _{\px^{\lambda}}$.
This is equivalent to the $2n+2 = N_r$ original equations (\ref{eq:eq_constr})
with $\bar{c}=\left\langle c\left(x\right)\right\rangle _{\px^{\lambda}}$.
Additionally, $\int dx\,\px^{\lambda}\left(x\right)=1$. The uniqueness
of the solution is guaranteed by $\left\langle c\left(x\right)\right\rangle _{\px^{\lambda}}$
being a strictly monotonically decreasing function of $\lambda$, which we will show in the extended version of the manuscript.
\end{IEEEproof}\vspace{0.4\baselineskip}

To conclude the proof of \emph{Step A}, we note that the same reasoning as in Case IIa 
applies.
The unique solution $\px^{\lambda^\ast}$ satisfies (\ref{eq:ineq_constr}) with equality at $x_{j}$,
$j=1,\ldots,N_r$. Moreover, \lemref{linear_i} shows that the l.h.s. of (\ref{eq:ineq_constr})
increases linearly between $x_{j}$ and $x_{j+1}$, and the r.h.s is concave due
to $\alpha\leq1$. Thus, (\ref{eq:ineq_constr}) is also satisfied
for all the points $x\in\left(x_{j},x_{j+1}\right)$, see \figref{cost} a). Hence, $\px^{\lambda^{\ast}}$
is the capacity-achieving input distribution $\px^{\ast}$ and its
support is $S_0$, i.e. that of the unconstrained case.


For \emph{Step B}, we first note that at $\bar{c}=\theta_{0}$, the mass $m_2$ vanishes (see Lemma \ref{lem:smallest-weight}) and $x_2$ is no longer in the support of 
$\px^{\lambda^{\ast}}$, see \figref{sketch_integer_case} (IIb),
which removes the first two difference equations in (\ref{eq:i_non_integer}).
We obtain (\ref{eq:m_hat_j_exlp-1}) and (\ref{eq:m_bar_j_expl-1}) 
but with the index $j$ startin at $j=3$.
These equations determine the relative weights within the set of masses
$M^{>}_{1}\coloneqq\left\{m_{j}\right\}_{j=3}^{N_r}$ as a function of $\lambda$.
The relative weight between $M_{1}^{<}\coloneqq\left\{m_{1}\right\}$ and $M^{>}_{1}$ can be
determined by the difference equation between the equality constraints for $x_1$ and $x_3$.
We then use \lemref{chain_composed} with $k=1$ to show the existence of a unique $\lambda^\ast$ such that
the cost constraint is met, and \lemref{ineq_at_x_2k} to show that the inequality constraint is also satisfied.
The masses in $M_{>}$ behave similarly to those in \emph{Step A} of the proof.
When $\bar{c}=\theta_1$, $m_4=0$ and one can apply the same reasoning as before setting $k=2$.
By showing that the relative size of the masses $M^{<}_{k}$ obeys (\ref{eq:masses_integer_case}),
we construct an iterative proof that is valid up to $\bar{c}\in(\theta_{n-1}, \theta_{n}]$.
\begin{lem}
\label{lem:chain_composed}(Equality constraint for $r\notin\mathbb{N}$ and $\bar{c}\in(\theta_{k},\theta_{k-1}]$)
For $\alpha\leq 1$, $r\notin\mathbb{N}$ and any given $\bar{c}\in(\theta_{k},\theta_{k-1}]$,
there is a unique solution to the equality constraint (\ref{eq:eq_constr})
given by the discrete probability distribution $\px^{\lambda^\ast}\left(x\right)=\sum_{\left\{ j\mid x_{j}\in S_{k}\right\} }m_{j}\left(\lambda^{\ast}\right)\,\delta\left(x-x_{j}\right)$. Here, $x_{j}$ is defined as in (\ref{eq:def_pos_unconstr}) and $m_{j}\left(\lambda\right)$, up to a normalization factor, is
given by (\ref{eq:masses_integer_case}) if $j<2k$, and (\ref{eq:def_masses_non_integer_case})
if $j>2k$.
\end{lem}
\begin{lem}
\label{lem:ineq_at_x_2k}(Inequality constraint for $r\notin\mathbb{N}$ and $\bar{c}\in(\theta_{k},\theta_{k-1}]$)
For $\alpha\leq 1$, $r\notin\mathbb{N}$ and any given
$\bar{c}\in(\theta_{k},\theta_{k-1}]$, the discrete probability distribution
$\px^{\lambda^\ast}\left(x\right)$ satisfies the inequality constraint (\ref{eq:ineq_constr}). Here, $x_{j}$ is defined as in (\ref{eq:def_pos_unconstr}) and $m_{j}\left(\lambda\right)$, up to a normalization factor, 
given by (\ref{eq:masses_integer_case}) if $j<2k$ and (\ref{eq:def_masses_non_integer_case})
if $j>2k$.
\end{lem}

Finally, for \emph{Step C}, we show that $m_{N_r}>0$ for $\bar{c}\in(0,\theta_{n-1}]$ so that the support remains $S_n$ in this interval.

This concludes the proof of IIb. 
\end{customproof}\vspace{0.4\baselineskip}
The capacity-achieving input distribution $\px^{\ast}\left(x\right)$
as a function of $r$ is depicted in \figref{capacity_achieving_input}.
Panel a) shows the unconstrained problem as discussed of Case Ia,  
 and panel b) depicts $\px^\ast$ with tight cost constraint as discussed in Cases IIa and IIb. 
\begin{customproof}{Case III}
First, we note that $\px^{\ast}\left(x\right)=0$ with $x\in\left[0,\epsilon\right]$,
$\epsilon>0$ is impossible because otherwise $\py^{\ast}\left(y\right)=0$
for $y\in\left[-b,-b+\epsilon\right]$ and hence, with 
\begin{equation}
    i\left(x;\px^{\ast}\right)=-\frac{1}{2b}\int_{x-b}^{x+b}dy\,\log\left[2b\,\py^{\ast}\left(y\right)\right],
\end{equation}
$i\left(x,\px^{\ast}\right)\rightarrow\infty$, which contradicts
(\ref{eq:ineq_constr}). For the same reason $\px^{\ast}=0$ on
the interval $\left[1-\epsilon,1\right]$, and gaps of width $d\geq2b$
in $S$ are incompatible with (\ref{eq:ineq_constr}). 

Now, we will prove by contradiction that $S$ cannot also have gaps
$g\coloneqq\left(x_{1},x_{2}\right)$ of finite measure, where $0<x_{1}<x_{2}<1$.
Assume that $x_{1},x_{2}\in S$ and $g\not \subseteq S$. Then, (\ref{eq:eq_constr})
has to be satisfied at and (\ref{eq:ineq_constr}) between the two
points $x_{1}$ and $x_{2}$. If $\alpha>1$, the r.h.s. of (\ref{eq:ineq_constr})
has a strictly convex shape, which, as we will show, cannot be matched by the l.h.s. of the equation. To this end, we move from $x_{1}$ to $x_{2}$ using the parametrization $x_\beta \coloneqq\left(1-\beta\right)x_{1}+\beta\, x_{2}$,
$\beta\in\left[0,1\right]$. Now, $i\left(x_\beta;\px^{\ast}\right)$ is defined as the integral of $f(y) \coloneqq -\frac{\log[2b\,\py^*(y)]}{2b}$ over $[x_\beta-b,x_\beta+b]$. We split this set into three subsets
$A_1 \coloneqq [x_\beta-b,x_2-b]$, 
$A_2 \coloneqq [x_2-b,x_1+b]$, and 
$A_3 \coloneqq [x_1+b, x_\beta+b]$. Note that $|A_1|=(1-\beta)(x_2-x_1)$ and $|A_3|=\beta(x_2-x_1)$. In addition, we define the left enlargement of $A_1$ as $A_1' = [x_1-b,x_2-b]$, with $|A_1'|=x_2-x_1$. Due to the gap, $\py^{\ast}\left(y\right)=\frac{1}{2b}\int_{y-b}^{y+b}dx\,\px^{\ast}\left(x\right)$
is a decreasing function of $y$ on the set $A_1'$, which implies that $f(y)$ is increasing and, due to the left enlargement of $A_1$, we have
\begin{equation}
    \frac{1}{|A_1'|} \int_{A_1'} dy \, f(y) \leq \frac{1}{|A_1|} \int_{A_1} dy \, f(y).
\end{equation}
Similarly, we can define the enlargement $A_3' = [x_1+b,x_2+b]$ of $A_3$ with $|A_3'|=x_2-x_1$. Since $f(y)$ is an decreasing function on $A_3'$ due to the gap, we obtain as before
\begin{equation}
    \frac{1}{|A_3'|} \int_{A_3'} dy\,f(y) \leq  \frac{1}{|A_3|} \int_{A_3} dy\,f(y) .
\end{equation}
Using the two inequalities above, we obtain
\begin{align}
    &i((1-\beta)\,x_1 + \beta\,x_2;\px^*) = \int_{A_1 \cup A_2 \cup A_3}dy\,f(y) \\
    \geq &(1-\beta) \int_{A_1'} dy\,f(y) + \int_{A_2} dy\,f(y) + \beta \int_{A_3'} dy\,f(y) \\
    = &(1-\beta)\, i(x_1;\px^*) + \beta \, i(x_2;\px^*),
\end{align}
using that $i(x_1;\px^*)$ and $i(x_2;\px^*)$ are the integrals of $f(y)$ over $A_1' \cup A_2$ and $A_2 \cup A_3'$, respectively.
This shows that $i\left(x,\px^{\ast}\right)$ is 
of concave shape, which contradicts (\ref{eq:ineq_constr}) due to the equalities at $x_1$ and $x_2$.
\end{customproof}
\vspace{0.4\baselineskip}
\figref{cost} b) shows the transition from discrete to full support of
$\px^\ast$ when $\alpha$ crosses 1.

\section{Discussion}
In this article, we computed the capacity-achieving input distribution for the uniform channel with peak amplitude constraint (PA) as well as expected cost constraint. We found two ways for the capacity-achieving input distribution to transition from discrete values to continuous values: either by increasing the cost function exponent $\alpha$ and crossing the critical exponent $\alpha^* \!=\! 1$ (provided that the cost constraint is active) or by decreasing the maximal cost $\bar{c}$ and crossing the critical cost $\bar{c}^*$ (provided that $\alpha>1$).

Remarkably, when the capacity-achieving input distribution is discrete, the possible position of the mass points cannot be at other locations than the ones given by $S_0$ (i.e. $S\subset S_0$) independently of the exponent $\alpha$ and the maximal cost $\bar{c}$ (even though the specific $S_k$ will depend on $\alpha$ and $\bar{c}$). This observation might hint towards a potentially simpler proof of the main theorem by using a generalization of the implicit function theorem. 

This study can be seen as an extension of the work of Oettli \cite{oettli_capacity-achieving_1974}, since we consider an additional (tunable) cost constraint which is the key ingredient that enables the phase transition between continuous and discrete capacity-achieving input distribution. This study also differs in two ways from the work of Tchamkerten \cite{tchamkerten_discreteness_2004}. First, we derive necessary and sufficient conditions (and not only sufficient conditions) for the emergence of discreteness for the capacity-achieving input distribution and secondly we consider an additive channel with bounded noise instead of unbounded noise. 

The present work could be extended in several directions. A first extension could be to remove the PA and replace it with a softer constraint (e.g. $c(x)\rightarrow x^\alpha + x^\beta, \forall x\leq 0$ and $\beta\geq 0$. The present PA corresponds to $\beta\rightarrow\infty$), whereas the absence of a PA would correspond to $\beta = 0$. This absence of PA could also be approached within the present framework in the limit of $\bar{c}\rightarrow 0$ and $r\rightarrow 0$. This extension, which would smoothly remove the PA, would help us to determine to what extent the hardness of the constraint leads to the discrete support of the capacity-achieving input distribution. %
Another extension could be to consider a generalization of the capacity problem in higher dimensions where the input is restricted to a $L_1$ ball, analogously to the $L_2$ ball constraint for the additive vector Gaussian channel \cite{shamai_capacity_1995,dytso_capacity_2019,eisen_capacity-achieving_2023}.

\section*{Acknowledgment}
This work has been supported by the Swiss National Science Foundation grant entitled "Why spikes?" (310030\_212247). 
\IEEEtriggeratref{8}

\bibliographystyle{IEEEtran}

\bibliography{UniformNoise}

\end{document}